\documentclass[sigconf]{acmart}

\usepackage{microtype}      % microtypography
\usepackage{xcolor}         % colors
\usepackage{amsmath}
\usepackage{setspace}
\usepackage{graphicx} 
\usepackage{float} 
\usepackage{subfigure} 
\usepackage{multirow}
\usepackage{diagbox}
\usepackage{color}
\usepackage{enumitem}
\usepackage{booktabs}
\usepackage{subfigure} 
\usepackage{cleveref}
\usepackage{tabularx}

\AtBeginDocument{%
  }

\copyrightyear{2024}
\acmYear{2024}
\setcopyright{acmlicensed}\acmConference[RecSys '24]{18th ACM Conference on Recommender Systems}{October 14--18, 2024}{Bari, Italy} \acmBooktitle{18th ACM Conference on Recommender Systems (RecSys '24), October 14--18, 2024, Bari, Italy}
\acmDOI{10.1145/3640457.3688164} 
\acmISBN{979-8-4007-0505-2/24/10}

\begin{document}

%%
%% The "title" command has an optional parameter,
%% allowing the author to define a "short title" to be used in page headers.
\title{Pay Attention to Attention for Sequential Recommendation}

%%
%% The "author" command and its associated commands are used to define
%% the authors and their affiliations.
%% Of note is the shared affiliation of the first two authors, and the
%% "authornote" and "authornotemark" commands
%% used to denote shared contribution to the research.
\author{Yuli Liu}
\affiliation{%
  \institution{Qinghai University \\
Intelligent Computing and Application Laboratory of Qinghai Province}
  \city{Xining 810016}
  \country{China}}
\email{liuyuli012@gmail.com}

\author{Min Liu}
\affiliation{%
  \institution{Qinghai University \\
 Intelligent Computing and Application  Laboratory of Qinghai Province}
  \city{Xining 810016}
  \country{China}}
\email{liumin5061@gmail.com}

\author{Xiaojing Liu}
\authornote{Corresponding author.}
\affiliation{%
  \institution{Qinghai University \\
 Intelligent Computing and Application Laboratory of Qinghai Province}
  \city{Xining 810016}
  \country{China}}
\email{liuxj@qhu.edu.cn}

%%
%% By default, the full list of authors will be used in the page
%% headers. Often, this list is too long, and will overlap
%% other information printed in the page headers. This command allows
%% the author to define a more concise list
%% of authors' names for this purpose.
\renewcommand{\shortauthors}{Yuli Liu, Min Liu, \& Xiaojing Liu}

%%
%% The abstract is a short summary of the work to be presented in the
%% article.
\begin{abstract}
  Transformer-based approaches have demonstrated remarkable success in various sequence-based tasks. However, traditional self-attention models may not sufficiently capture the intricate dependencies within items in sequential recommendation scenarios. This is due to the lack of explicit emphasis on attention weights, which play a critical role in allocating attention and understanding item-to-item correlations. To better exploit the potential of attention weights and improve the capability of \textbf{s}equential \textbf{r}ecommendation in learning high-order dependencies, we propose a novel \textbf{s}equential \textbf{r}ecommendation (SR) approach called \textbf{a}ttention \textbf{w}eight \textbf{r}efinement  (AWRSR). AWRSR enhances the effectiveness of self-attention by additionally paying attention to attention weights, allowing for more refined attention distributions of correlations among items. 
%This approach helps capture higher-order dependencies and further amplifies the crucial information present in the attention weights.
We conduct comprehensive experiments on multiple real-world datasets, demonstrating that our approach consistently outperforms state-of-the-art SR models. Moreover, we provide a thorough analysis of AWRSR's effectiveness in capturing higher-level dependencies. These findings suggest that AWRSR offers a promising new direction for enhancing the performance of self-attention architecture in SR tasks, with potential applications in other sequence-based problems as well.
\end{abstract}

%%
%% The code below is generated by the tool at http://dl.acm.org/ccs.cfm.
%% Please copy and paste the code instead of the example below.
%%



%%
%% Keywords. The author(s) should pick words that accurately describe
%% the work being presented. Separate the keywords with commas.
%% A "teaser" image appears between the author and affiliation
%% information and the body of the document, and typically spans the
%% page.

%%
%% This command processes the author and affiliation and title
%% information and builds the first part of the formatted document.
\maketitle

\section{Introduction}
%learning higher-order dependencies
%refine representations by transforming attention weights calculated from the original representations.
%context-aware
%the attention weights themselves are also being transformed
Sequential Recommendation \cite{liu2016context, kang2018self} (SR) aims to predict a user's next interaction by considering their historical behavior, making it a practical and advanced variant of traditional recommendation systems \cite{sarwar2001item, pazzani2007content,he2017neural, schafer2007collaborative, liu2024learning}. %This task closely mirrors real-world user-item interactions, where actions are influenced by a series of preceding events rather than isolated incidents. As such, sequential recommendation systems provide a more accurate and contextually relevant reflection of user behavior, compared to their traditional counterparts (such as content-based recommendation \cite{sarwar2001item, pazzani2007content} and collaborative filtering \cite{he2017neural, schafer2007collaborative}). 
A key challenge in developing effective SR systems is the need to model sequential dependencies within a user's interaction history. In order to capture the sequential dependencies, the construction of a recommendation framework based on sequence models is primarily employed. 
Among various sequence models, the Self-Attention (SA) method \cite{vaswani2017attention, kang2018self, he2016fusing, li2017image, zhou2022jump}, which forms the core of the transformer architecture, has emerged as a particularly successful approach in the field of SR. Compared to traditional models like Recurrent Neural Networks (RNNs) \cite{donkers2017sequential, medsker2001recurrent, cho2014learning, devooght2017long} and Markov chains \cite{he2016fusing, rendle2010factorizing}, SA exhibits unique capabilities in modeling complex temporal dependencies within a user's item interaction sequence.  
The success of self-attention can be primarily attributed to its attention weights, which measure the similarity or correlation between any two elements in a sequence, and thus adaptively adjust the attention distribution. The most common approach to calculate attention weights is the dot product, where the weights are computed by taking the dot product between a query vector and key vectors \cite{wang2018attention, zhang2018next, abnar2020quantifying}. To effectively measure attention, the use of scaled Wasserstein distances between items' representations has been proposed \cite{fan2022sequential}. To better model complex patterns and capture intricate dependencies, considering higher-order item-item transitions has also been an active research topic. There has been a range of work exploring higher-order relationships in SR systems. Early approaches utilize higher-order Markov chains \cite{rendle2010factorizing} and RNNs \cite{hidasi2015session} to capture complex temporal dynamics. Recently, leveraging graph-based techniques such as Graph Convolutional Networks (GCNs) \cite{ying2018graph} and graph variational method \cite{ding2021semi} to model higher-order connectivity has been introduced. 

These studies highlight the importance of attention weights and higher-order relationships in learning expressive representations for recommendation systems. Despite the importance being highlighted, it is surprising that existing literature has not yet paid attention to attention weights with the motivation of capturing higher-order transitions by exploring correlations within attention weights. This naturally leads us to pose an intuitive question: Given the remarkable ability of attention weights to allocate attention distributions, which stems from the ability of measuring the correlations of each item with all other items in the sequence, it prompts the question of whether we can further refine higher-order correlations by paying attention to these attention weights? In essence, this question boils down to: why do we not utilize attention weights as elevated representations (item's correlations with others) to model the higher-order dependencies? 

Motivated by this question, we propose a novel self-attention architecture for Sequential Recommendation called Attention Weight Refinement (AWRSR). AWRSR refines attention weights by specifically focusing on examining correlations among the attention weights themselves, enabling our approach to learn higher-order relationships between input features and attention weights. 
Specifically, instead of using these weights as in traditional self-attention architecture that multiplies the attention matrix by values, we attempt to transform the attention weights into a new space (\textit{i.e.}, correlation representations) using learnable matrices. 
These refined weights are then used to compute higher-order attention weights. In other words, we pay attention to the original attention weights.
%This refined attention mechanism is designed to efficiently capture higher-order correlations by leveraging the valuable information hidden within the attention weight matrices, offering the potential for improved performance in sequential recommendation tasks. 

The main contributions of this work are:
(i) We highlight the limitations of traditional self-attention mechanisms in sequential recommendation, specifically their inability to explicitly exploit higher-order correlations within attention weights. This serves as our primary motivation; (ii) Driven by this motivation, we propose a novel self-attention architecture that refines attention weights by paying attention to attention weight correlations for capturing higher-order dependencies and uncovering valuable information hidden within the weights; (iii) Several refinement mechanisms for the attention weights are designed. A series of heuristic experiments performed within state-of-the-art and representative attentive approaches substantiate the benefits of our self-attention architecture. 
    %\item Through detailed comparative analyses, we offer a deep understanding of why our refinement self-attention mechanisms outperform existing methods.
This is a crucial area for future research as understanding the interplay among attention weights can potentially lead to more robust and efficient transformer models.

\vspace{-1mm}
\section{AWRSR Framewrok}
\vspace{-0.5mm}
Before delving into our method, we first introduce the problem definition of SR and then explore a variety of refinement mechanisms. 

\vspace{-1mm}
\subsection{Problem Definition}
\vspace{-0.5mm}

{\begin{figure}
  \centering
  \includegraphics[width=.9\linewidth]{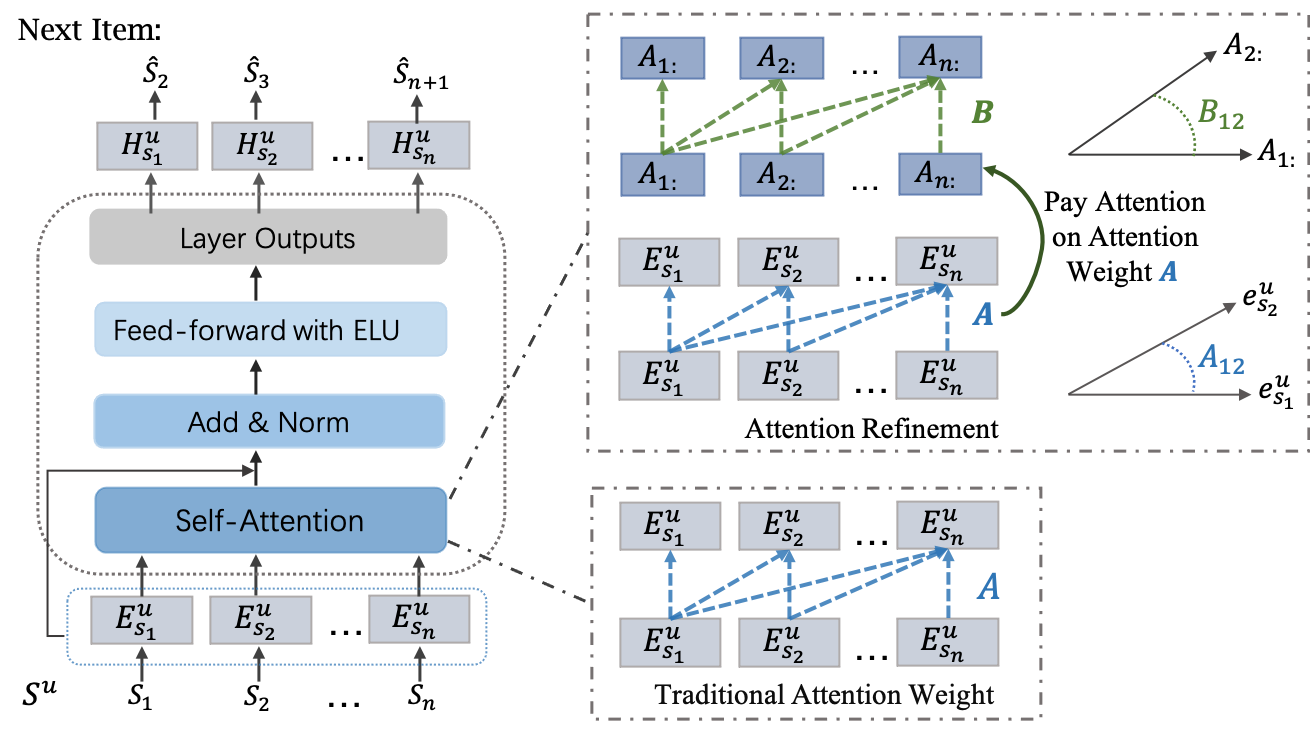}
  %\fbox{\rule[-.5cm]{0cm}{4cm} \rule[-.5cm]{4cm}{0cm}}
  \vspace{-2mm}
  \caption{Attention weight refinement for SR.}
  \vspace{-3mm}
  \label{img-architecture}
\end{figure}

In the context of a user set $\mathcal{U}$ and an item set $\mathcal{V}$, we can represent the interactions of each user $u \in \mathcal{U}$ in a chronological sequence $\mathcal{S}^u=[s_1, s_2, ..., s_{|\mathcal{S}^u|}]$, with $s_i \in \mathcal{V}$ as the $i$-th item that user $u$ interacted with. The objective of SR is to generate a top-N list of items as potential next interactions, given previous sequence. 

\vspace{-1mm}
\subsection{Self-attention based SR Models}
\label{sec:self-attention}
\vspace{-0.5mm}
%As we aim to capture the intricate dependencies in sequential recommendation by refining higher-order transitions from self-attention weights, our discussion starts with the foundational self-attention-based sequential recommendation method.
%Following that, we will delve into two distinct yet representative methodologies for computing attention weights. These methodologies serve as the basis for our proposed refinements. The self-attention mechanism intuitively assigns different attention weights to interrelated items in a sequence, reflecting their varying influence based on their positions. The architecture of self-attention based sequential recommendation is as depicted in the left segment of \Cref{img-architecture}. 

%Specifically, given an action sequence $\mathcal{S}^u$ of user $u$ and a maximum sequence length $n$, the sequence is usually standardized by truncating or padding with zeros. Combining the defined item embedding matrix $\mathbf{M} \in \mathbb{R}^{|\mathcal{V}| \times d}$ (where $d$ represents the latent dimensions) and trainable positional embedding $\mathbf{P} \in \mathbb{R}^{n \times d}$, the sequence embedding matrix can be created,
%\begin{equation}
%\hat{\mathbf{E}}_{\mathcal{S}^u}=\left[\mathbf{m}_{s_1}+\mathbf{p}_{s_1}, \mathbf{m}_{s_2}+\mathbf{p}_{s_2}, \ldots, \mathbf{m}_{s_n}+\mathbf{p}_{s_n}\right].
%\end{equation}

The most common approach in self-attention architecture is to use scaled dot-product operation among items within a sequence to deduce their interrelations,
\begin{equation}
\mathrm{SA}(\mathbf{Q}, \mathbf{K}, \mathbf{V})=\operatorname{softmax}\left(\frac{\mathbf{Q K ^ { \top }}}{\sqrt{d}}\right) \mathbf{V},
\label{equ-SA}
\end{equation}
where $\mathbf{Q}=\hat{\mathbf{E}}_{\mathcal{S}^u} \mathbf{W}^Q, \mathbf{K}=\hat{\mathbf{E}}_{\mathcal{S}^u} \mathbf{W}^K$, and $\mathbf{V}=\hat{\mathbf{E}}_{\mathcal{S}^u} \mathbf{W}^V$, representing the queries, keys, and values, respectively. 
%This effective and common attention weight operation (dot product) is widely used not only in the field of sequential recommendation (including the seminal work SASRec \cite{kang2018self} and other recommendation models \cite{zhang2019feature, li2021lightweight}), but also in multiple tasks such as neural language \cite{zhou2022jump} and image caption \cite{li2017image， zhou2022jump}. 
The attention value $\mathbf{A}_{k t}$ between item $s_k$ and item $s_t$ in $k$-th and $t$-th positions of the sequence can be calculated as
$\mathbf{A}_{k t}=\mathbf{Q}_k \mathbf{K}_t^{\top} / \sqrt{d}$. $\mathbf{A} \in$ $\mathbb{R}^{n \times n}$ denotes the SA weights. 

Recently, a powerful and effective benchmark STOSA \cite{fan2022sequential} for calculating attention weights has emerged, which adopts Wasserstein distance \cite{ruschendorf1985wasserstein} to measure the discrepancy between stochastic embeddings of two items with the motivation of satisfying the triangle inequality \cite{clement2008elementary}. 
Formally, given two items $s_k$ and $s_t$, the corresponding stochastic embeddings are two distributions $\mathcal{N}\left(\mu_{k}, \Sigma_{k}\right)$ and $\mathcal{N}\left(\mu_{t}, \Sigma_{t}\right)$,
where $\mu_{k}=\hat{\mathbf{E}}_{k}^\mu \textbf{W}_\mu, \Sigma=\operatorname{ELU}\left(\operatorname{diag}\left(\hat{\mathbf{E}}_{k}^{\Sigma} \textbf{W}_{\Sigma}\right)\right)+1$ ($k$ and $t$ are used for brevity). Exponential Linear Unit (ELU) is used to guarantee the positive definite property of covariance. %As claimed in STOSA, the dimensions of $\hat{\mathbf{E}}_{k}^\mu$ and $\hat{\mathbf{E}}_{k}^\Sigma$ are set to $d/2$ to ensure a fair comparison with SASRec. 
STOSA defines the attention weight as the negative 2-Wasserstein distance.
%\begin{equation}
%\begin{aligned}
%\mathbf{A}_{k t} =-\left(&\left\|\mu_{s_k}-%\mu_{s_t}\right\|_2^2 \\
%&+\operatorname{trace}\left(\Sigma_{s_k}+\Sigma_{s_t}-2\left(\Sigma_{s_k}^{1 / 2} \Sigma_{s_k} \Sigma_{s_t}^{1 / 2}\right)^{1 / 2}\right)\right).
%\label{equ_distance}
%\end{aligned}
%\end{equation}

Existing attention weight operation generally revolves around measuring the similarity or distance between items' embeddings or distributions, as depicted in the lower right corner segment of \Cref{img-architecture}. They fail to consider the potential correlations within attention weights that could further uncover higher-order transitions within the attention weights themselves. 
As shown in the segment located in the upper right corner, the attention weight matrix \textbf{A}, derived from the representations of items in a sequence, encodes the relationships of each item in the sequence with respect to all other items. For example, $\mathbf{A}_{1:} \in \mathbb{R}^n$ signifies the allocation of attention or importance from the first item to all other items in the sequence. 
This implies that refining these weights, which inherently capture attention correlations, holds potential for modeling higher-order transitions. Given this context, comparing $\mathbf{A}_{1:}$ and $\mathbf{A}_{2:}$ can help to deepen our understanding of the dependencies between two items (in the first and second positions) of the sequence. 

In this work, the primary objective is to refine attention weights. To achieve this, we draw from two contrasting but notable self-attention architectures, \textit{i.e.}, previously mentioned SASRec \cite{kang2018self} and STOSA \cite{fan2022sequential}  (each yielding distinct interpretations of attention weights). 
%Building upon these unique attention weights, we then develop diverse refinement mechanisms, with the intention of optimizing these weights to improve performance in sequential recommendation. 
The choice of selecting these two models is primarily motivated by their individual characteristics and contributions. SASRec is a seminal self-attention-based model for sequential recommendation, inspiring numerous subsequent studies \cite{li2021lightweight, sun2019bert4rec}, and STOSA calculates attention weights from a fresh and distinctly different perspective (\textit{i.e.}, discrepancy of probabilistic distributions). If our refinement mechanisms improve upon SASRec, it indicates the general adaptability of our approach in enhancing a well-established, traditional SA model, signifying the potential of our method to be universally applicable across similar standard models. 
On the other hand, their effectiveness on STOSA will signify the flexibility and robustness of our method. 
%Additionally, our mechanisms (deployed on both SASRec and STOSA) incorporate other components from the Transformer architecture as utilized in SASRec, including the point-wise feed-forward network, the residual connection, and layer normalization. 

\vspace{-1mm}
\subsection{Refinement }
\vspace{-0.5mm}

We now introduce our "Pay Attention to Attention" refinement mechanisms devised for two distinct architectures (SASRec and STOSA) that each yields a attention weight matrix, collectively denoted as $\mathbf{A} \in \mathbb{R}^{n \times n}$. $n$
represents the maximum length. 
%If the sequence length is greater than $n$, the most recent $n$ actions will be considered. If the sequence length is less than $n$, a ‘padding’ item will be repeatedly add to the left until the length is $n$. 
Our proposed mechanisms operate directly on $\mathbf{A}$. 

\textbf{Mechanism 1:} Simple refinement ($\boldsymbol{M}_{simp}$). This mechanism simply applies a new set of trainable matrices to $\textbf{A}$, transforming it into new queries and keys for computing higher-level weights: 
\begin{equation}
    \mathbf{B}_{kt} = \frac{\left(\mathbf{A}_{k:} \mathbf{W}^{(RQ)}\right)^{\top} \left({\mathbf{A}_{t:} \mathbf{W}^{(RK)}} \right)}{\sqrt{d}},
\end{equation} 
where $\mathbf{W}^{(RQ)} \in \mathbb{R}^{n \times n}$ and $\mathbf{W}^{(RK)} \in \mathbb{R}^{n \times n}$ denote the  trainable matrices. 
We then perform $\operatorname{softmax}(\cdot)$ function on the new weight matrix $\mathbf{B}\in \mathbb{R}^{n \times n}$. The multiplication of $\operatorname{softmax}(\mathbf{B})$ with $\mathbf{V}$ corresponds to summing the values, which means that each value is scaled by the refined higher-order attention weight.

\textbf{Mechanism 2:} Value-weighted refinement ($\boldsymbol{M}_{value}$) further transforms the new weights (\textit{i.e.}, $\mathbf{B}$ from $\boldsymbol{M}_{simp}$) by attention-determined values ($\mathbf{W}^{(RV)}\mathbf{A}$, where $\mathbf{W}^{(RV)} \in \mathbb{R}^{n \times n}$), hoping to distill and re-express the weight correlations in a more sophisticated, higher-order space. We give the calculation of matrix form for simplication,
\begin{equation}
\mathbf{B} = \mathbf{W}^{(RV)}\mathbf{A} \operatorname{softmax} \left(\left(\mathbf{W}^{(RK)} \mathbf{A}\right)^{\top} \mathbf{W}^{(RQ)} \mathbf{A} \right).
\end{equation}  

\textbf{Mechanism 3:} Additive refinement ($\boldsymbol{M}_{add}$) is proposed with the intention of incorporating / balancing attention weights of different levels, calculated by  
\begin{equation}
    \mathbf{B}_{kt} = \left(\frac{\left(\mathbf{A}_{k:} \mathbf{W}^{(RK)}\right)^{\top} \left({\mathbf{A}_{t:} \mathbf{W}^{(RQ)}} \right)}{\sqrt{d}} + \mathbf{A}_{kt}\right)/2.
\end{equation}

\textbf{Mechanism 4:} Stochastic refinement
($\boldsymbol{M}_{stoc}$) is tailored to stochastic SA architecture STOSA. We try to transform the original stochastic weight $\textbf{A}$ of STOSA into a new form that potentially maintains the probabilistic nature of the weights. We calculate a new attention distribution $\mathcal{N}\left(\mu^{R}_k, \Sigma^{R}_k\right)$ for item $s_k$, where $\mu^{R}_k=\mathbf{A}_{k:} \mathbf{W}_{\mu}^{(R)}, \Sigma^{R}_k=\operatorname{ELU}\left(\operatorname{diag}\left(\mathbf{A}_{k:} \mathbf{W}_{\Sigma}^{(R)}\right)\right)+1$, where $\mathbf{W}_\mu^{(R)} \in \mathbb{R}^{n \times n} $ and $\mathbf{W}_{\Sigma}^{(R)} \in \mathbb{R}^{n \times n}$ are two stochastic matrices. %Then the new stochastic aware weight matrix $\textbf{B}$ can be calculated referring to \Cref{equ_distance}.

\vspace{-1mm}
\section{Experiments}
\vspace{-0.5mm}
In this section, we validate and explain the effectiveness of the proposed AWRSR in multiple aspects. 

\subsection{Experimental Settings}

\begin{table}
\centering
  \fontsize{8.2}{9.9}\selectfont
  \caption{Statistics of the datasets.}
  \vspace{-1.2mm}
  \setlength{\tabcolsep}{4.8mm}{
    \begin{tabular}{ccccc}
    \hline
    \ \ \ Dataset \ &\#Users&\#Items&\#Interactions \ \ \ \\
    \hline
    \textbf{Beauty}& 52.0k & 57.2k & 0.4M \\
    \textbf{ML-M}& 6.0k & 3.4k & 1.0M \\
    \textbf{Anime}& 73.5k & 12.2k & 1.0M \\
    \hline
    \end{tabular}}
    \label{tab-dataset}
    \vspace{-2.1mm}
\end{table}

\textbf{Datasets.} Three widely used public benchmark datasets are selected for evaluation:  
(\textit{i}) We employ the 1M version (\textbf{ML-1M}) of the MovieLens dataset, which is commonly used and trusted in the field of recommendation; (\textit{ii}) The popular dataset \textbf{Anime} that consists of 1 million ratings from users to anime has also been selected in this work;
(\textit{iii}) Amazon dataset, known for its high sparsity, has been widely used in recommendation experiments. We leverage the \textbf{Beauty} product category for our analysis.
%We treat the presence of ratings as positive interactions. 
Following \cite{fan2022sequential, kang2018self}, the 5-core setting is also adopted by filtering out users with less than 5 interactions. 
%We arrange the user interactions in the datasets based on the timestamps of each rating, forming a chronological sequence. 
The most recent interaction is set aside for testing, the second last one for validation, and all other interactions for training. 
Details of datasets' statistics are summarized in \Cref{tab-dataset}. \\
\textbf{Compared Methods.} As mentioned above, we deploy our refinement mechanisms in the context of STOSA \cite{fan2022sequential} and SASRec \cite{kang2018self}, creating the new sequential recommendation framework AWRSR. If the reworked models exhibit improved performance compared to the original methods, it validates the effectiveness of our proposed mechanisms. Therefore, in the subsequent performance comparisons, we will assess the performance of the reworked models with our proposed mechanisms against the specific original model. \\
\textbf{Evaluation.} %For each user, all items' prediction scores calculated by STOSA-based and SASRec-based models are from the 2-Wasserstein distance (a smaller distance score indicates a higher probability of the next item) and the MF layer (dot-product method), respectively, which are sorted in ascending order to generate the top- $\mathrm{N}$ recommendation list. 
Two standard top-$\mathrm{N}$ ranking evaluation metrics, Recall@N (Re@N) and NDCG@N (Nd@N) are applied. We report the performance when $N=1$ and $N=5$ for the overall comparisons.  \\
\textbf{Implementation Details.} We implement our mechanisms based on the code released by STOSA %\footnote{\url{https://github.com/zfan20/STOSA}} 
and SASRec %\footnote{\url{https://github.com/RUCAIBox/CIKM2020-S3Rec}} 
with Pytorch in a Tesla K40 GPU. We first grid search all parameters on the original model, and then deploy our mechanisms on the original model with the tuned hyper-parameters. This means that experiments of the reworked models and the corresponding original model are definitely performed under the same configurations for a fair comparison. The test performance based on the best validation results are reported.
For two original models, we search the embedding dimension in $\{64,128\}$. As STOSA has both mean and covariance embeddings, only $\{32,64\}$ is searched. We also search max sequence length from $\{20, 30, 50, 100\}$. We tune the learning rate in $\left\{10^{-3}, 10^{-4}\right\}$, and search the $L_2$ regularization weight from $\left\{10^{-1}, 10^{-2}, 10^{-3}\right\}$. The dropout rate is searched from $\{0.3,0.5,0.7\}$. 
For sequential models, we search number of heads and number of layers both in $\{1,2,3,4\}$. 
We adopt the early stopping strategy that model optimization stops when the validation NDCG@5 does not increase for 20 epochs. 
%%We promise to release the implementations (codes) and pre-processed data upon publication. 

\vspace{-1mm}
\subsection{Overall Comparisons}
\vspace{-0.5mm}

\begin{table*}[tp]
\centering
  \fontsize{8.5}{10.5}\selectfont
  \caption{Overall performance comparison. The best results are in bold. The improvements (\%) of AWRSR the original method are denoted by the corresponding subscripts. Improvements over baselines are statistically significant with $p<0.01$.}
  \vspace{-1.5mm}
  \setlength{\tabcolsep}{1.5mm}{
  \label{tab:performance_comparison}
    \begin{tabular}{c|c|ccc|ccc|ccc}
    \toprule[1pt]
    \multirow{2}{*}{ }&\multirow{2}{*}{Method}&\multicolumn{3}{c|}{\textbf{Beauty}}&\multicolumn{3}{c|}{\textbf{ML-1M}}&\multicolumn{3}{c}{\textbf{Anime}}\\  &&Recall@1&Recall@5&NDCG@5&Recall@1&Recall@5&NDCG@5&Recall@1&Recall@5&NDCG@5  \cr \hline
    \multirow{5}{*}[2pt]{SASRec} 
    &Original & 0.0097&0.0371 & 0.0234& 0.0265 &0.0959 &0.0616 &0.1209&0.2978 &0.2112\\ \cline{2-11}
    &$\boldsymbol{M}_{simp}$ &0.0104$_{+7.22}$ &0.0376$_{+1.35}$ &0.0243$_{+3.85}$ & \textbf{0.0291}$_{+9.81}$ &0.1004$_{+4.69}$ &0.0649$_{+5.35}$ &\textbf{0.1237}$_{+2.31}$ &\textbf{0.3247}$_{+9.04}$ & 0.2274$_{+7.67}$\\
    &$\boldsymbol{M}_{value}$ &\textbf{0.0116}$_{+19.59}$ &0.0376$_{+1.35}$ &\textbf{0.0251}$_{+7.26}$ & 0.0267$_{+0.75}$ &0.0976$_{+1.78}$ &0.0642$_{+4.22}$ &0.1269$_{4.96}$ &0.3083 $_{+3.53}$&0.2165$_{+2.50}$\\
    &$\boldsymbol{M}_{add}$ &0.0108$_{+11.34}$ &\textbf{0.0380}$_{+2.43}$ &0.0239$_{+2.14}$ & 0.0289$_{+9.06}$ &\textbf{0.1013}$_{+5.63}$ &\textbf{0.0654}$_{+6.17}$ &0.1211$_{+0.25}$ &0.3225 $_{+8.31}$&\textbf{0.2290}$_{+8.42}$\\ 
     \midrule[0.7pt]
    \multirow{5}{*}[0pt]{STOSA} 
    &Original & 0.0150 &0.0422 &0.0285 &0.0218 &0.0720 &0.0475 &0.1101 &0.2716 &0.1934 \\ \cline{2-11}
&$\boldsymbol{M}_{simp}$& 0.0153$_{+2.00}$ &\textbf{0.0453}$_{+7.35}$ &0.0304$_{+6.67}$ &0.0221$_{+1.38}$ &0.0756$_{+5.00}$ &0.0489$_{+2.95}$ &0.1220$_{+10.81}$ &0.3018$_{+11.11}$ &0.2143$_{+10.84}$\\
&$\boldsymbol{M}_{value}$ & 0.0154$_{+2.67}$ &0.0428$_{+1.42}$ &0.0286$_{+0.35}$ &\textbf{0.0246}$_{+12.84}$ &\textbf{0.0789}$_{+9.58}$ &\textbf{0.0516}$_{+8.63}$ &\textbf{0.1260}$_{+14.44}$ &\textbf{0.3029}$_{+11.50}$ &\textbf{0.2171}$_{+12.27}$\\
&$\boldsymbol{M}_{add}$ &\textbf{0.0162}$_{+8.00}$ &0.0442$_{+4.74}$ &\textbf{0.0305}$_{+7.02}$ &0.0223$_{+2.29}$ &0.0750$_{+4.17}$ &0.0481$_{+1.26}$ &0.1222$_{+11.00}$ &0.2987$_{+9.96}$ &0.2147$_{+11.01}$\\
&$\boldsymbol{M}_{stoc}$ &0.0154$_{+2.67}$ &0.0425$_{+0.71}$ &0.0276$_{-3.16}$ &0.0225$_{+3.21}$ &0.0733$_{+1.81}$ &0.0478$_{+0.63}$ &0.1194$_{+8.45}$ &0.2878$_{+5.96}$ &0.2073$_{+7.19}$\\
    \bottomrule[1pt]
    \end{tabular}}
    \label{tab:overall-performance}
    \vspace{-1.5mm}
\end{table*}

By evaluating the performance of all models as shown in \Cref{tab:overall-performance}, the following observations are made:
%The reworked models that integrate our proposed mechanisms within STOSA and SASRec, have all led to performance improvements across all three datasets. 
(\romannumeral1) Our straightforward mechanism $\boldsymbol{M}_{simp}$ that simply transforms the original attention matrices (from SASRec and STOSA) by new query weights and key weights for calculating the correlations from items' attention distributions finally achieves performance improvements compared to original models. Even this straightforward refinement can significantly boost the self-attention approach's capacity, suggesting the potential existence of correlations among attention weights;
Refining these correlations enhances the ability of self-attention approach to model higher-order  transitions.
(\romannumeral2) Among multiple refinement mechanisms (three types on SASRec and four on STOSA), the simple and intuitive mechanism $\boldsymbol{M}_{simp}$ consistently achieves improvements across different datasets and original models. This indicates the stability and applicability of this approach in refining higher-order correlations for sequential recommendation;
%\item Of all the implemented mechanisms (across three datasets and two models), the most significant improvements are achieved by the $\boldsymbol{M}_{value}$ mechanism, which also exhibits the greatest variability in performance. The effectiveness of $\boldsymbol{M}_{value}$ is related to the original models' performance, as it further transforms their attention weights into values. If the potential of the original model has not been fully exploited, this further higher-level transformation can provide the model with opportunities to capture new correlations, thereby largely enhancing the final results. For instance, in the beauty dataset experiments, SASRec's performance is inferior to that of STOSA. However, once $\boldsymbol{M}_{value}$ is deployed, SASRec's performance finds substantial room for improvement. A similar conclusion can be found on ML-1M.
(\romannumeral3) In contrast to $\boldsymbol{M}_{value}$,  $\boldsymbol{M}_{add}$ mechanism sometimes shows less noticeable enhancements. However, it generally outperforms both the $\boldsymbol{M}_{simp}$ and $\boldsymbol{M}_{value}$. This can be attributed to the design of $\boldsymbol{M}_{add}$, which merges higher-order correlations and inherent correlations among original weights, thereby achieving a certain balance; 
(\romannumeral4) The specifically designed $\boldsymbol{M}_{stoc}$ mechanism for STOSA, does not achieve significant performance improvements. This may be due to the inherent challenge of effectively transforming attention weights into a promising stochastic distribution.

To more comprehensively demonstrate the effectiveness of our mechanisms, we use \Cref{fig:NDCG-K} to compare the NDCG performance of our mechanisms and traditional models under different Top-N recommendations on Beauty. As observed, our proposed mechanisms (especially $\boldsymbol{M}_{simp}$ and $\boldsymbol{M}_{add}$) consistently outperform original models across various lengths of recommendation lists. This implies that the refinement mechanisms can better understand and leverage complex relationships among items by capturing higher-order and long-range dependencies within sequences.

\begin{figure}
\centering
    \subfigure[SASRec on Beauty]{
    \includegraphics[width=0.48\linewidth]{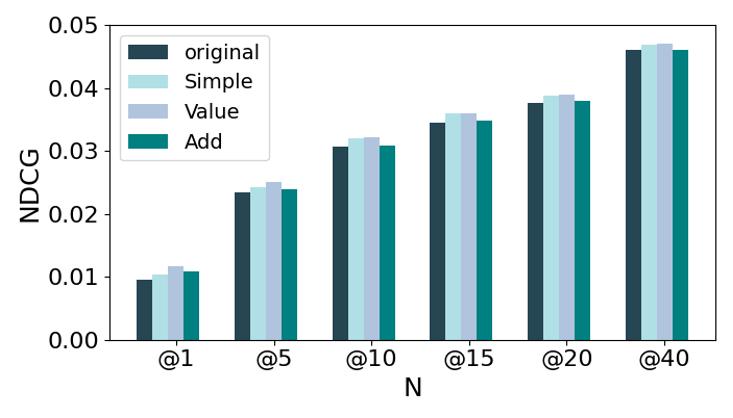}}
    \subfigure[STOSA on Beauty]{
    \includegraphics[width=0.48\linewidth]{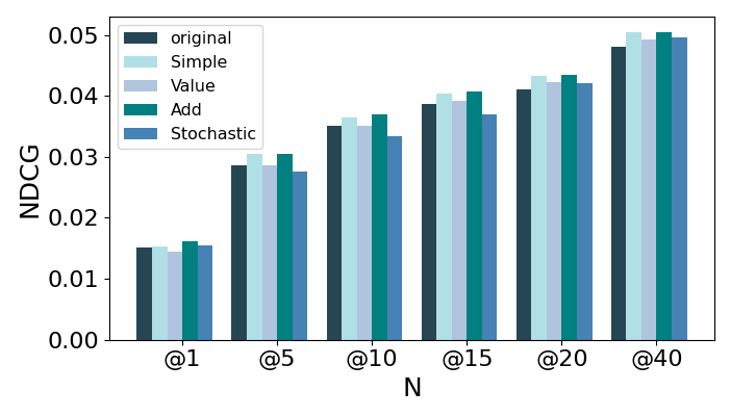}}
    \subfigure[SASRec on ML-1M]{
    \includegraphics[width=0.48\linewidth]{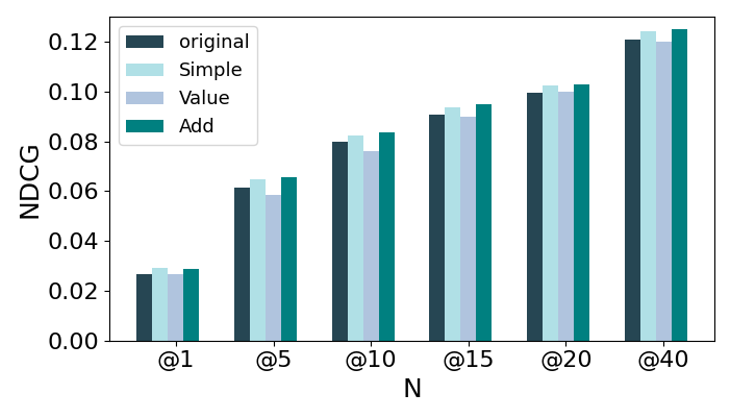}}
    \subfigure[STOSA on ML-1M]{
    \includegraphics[width=0.48\linewidth]{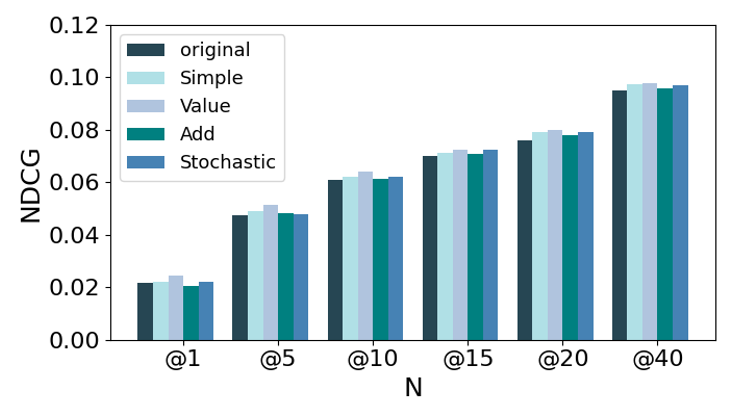}}
    \subfigure[SASRec on Anime]{
    \includegraphics[width=0.48\linewidth]{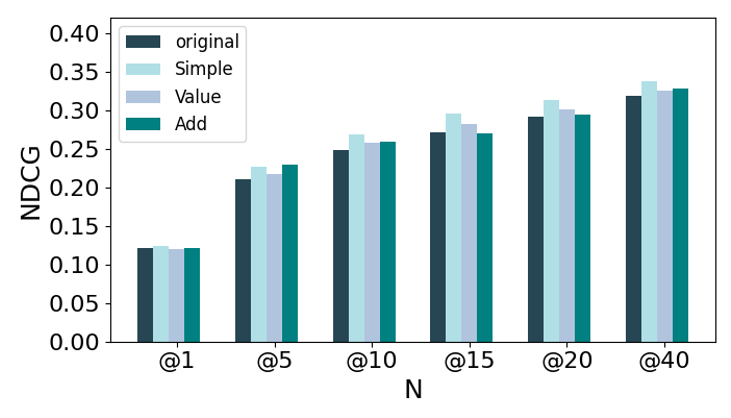}}
    \subfigure[STOSA on Anime]{
    \includegraphics[width=0.48\linewidth]{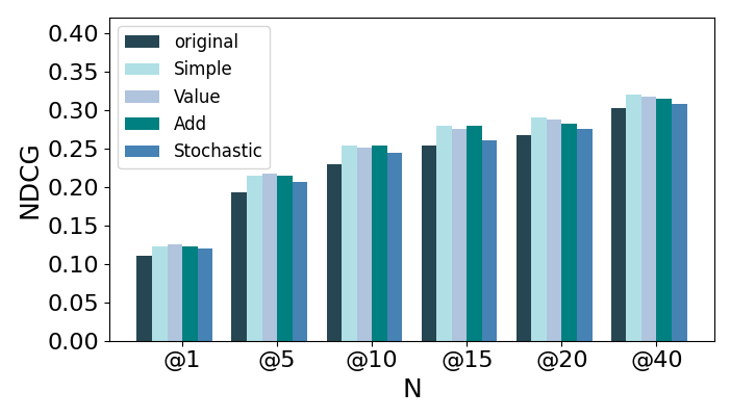}}
    \vspace{-2mm}
\caption{The NDCG performance \textit{w.r.t.} different Top@N 
%on Beauty using SASRec and STOSA
.}
\label{fig:NDCG-K}
\vspace{-3mm}
\end{figure}

Our "Pay Attention to Attention" methodology stands out from the traditional SA schemes (\textit{e.g.}, multi-layer SA and multi-head SA). Multi-head SA involves utilizing parallel attention heads to capture different aspects of the input. Multi-layer SA focuses on stacking repeated layers, and attention weights are used to obtain new representations.
In contrast, our mechanism dives into the structure of attention weights themselves to extract higher-level correlations from the attention distributions among items in a sequence. This not only enhances our understanding of the attention structure, but also achieves superior performance without incurring the same level of computational costs as the multi-layer self-attention (detailed in \Cref{sec-complexity}).

\subsection{Experimental Results Analysis}

\begin{figure}
\centering
     \subfigure[ STOSA]{
    \includegraphics[width=0.315\linewidth]{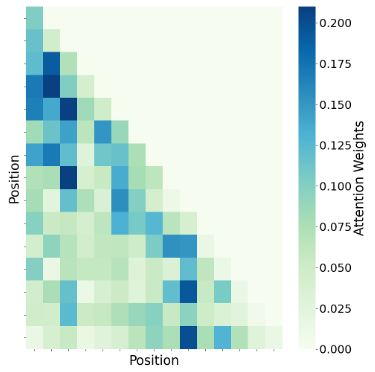}}
    \subfigure[ Simple-A]{
    \includegraphics[width=0.315\linewidth]{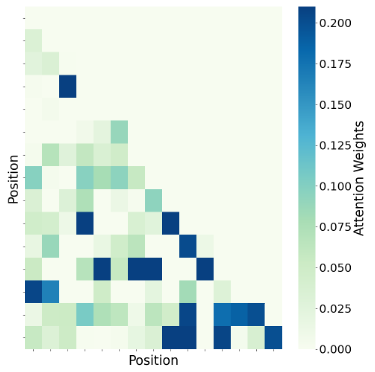}}
    \subfigure[ Simple-B]{
    \includegraphics[width=0.315\linewidth]{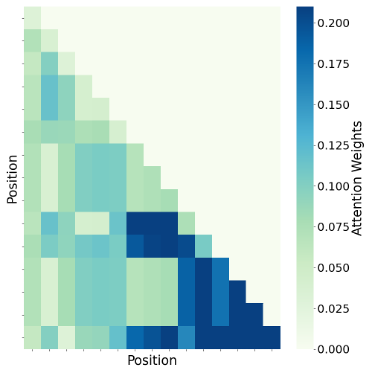}}
    \subfigure[SASRec]{
    \includegraphics[width=0.315\linewidth]{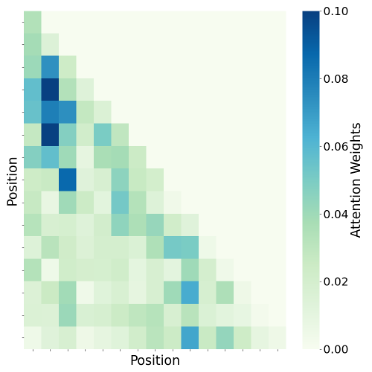}}
    \subfigure[Simple-A]{
    \includegraphics[width=0.315\linewidth]{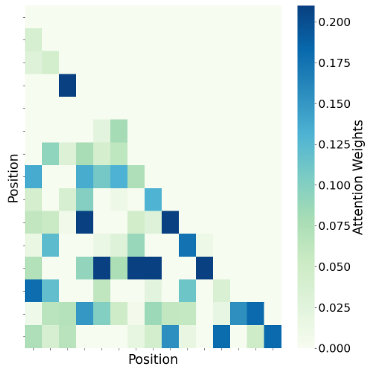}}
    \subfigure[Simple-B]{
    \includegraphics[width=0.315\linewidth]{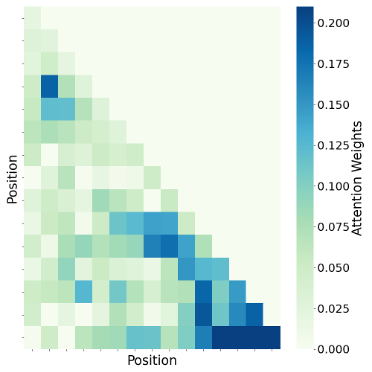}}
 \vspace{-2mm}
\caption{Attention weights visualizations.}
\vspace{-3mm}
\label{fig:attention-weights}
\end{figure}

\Cref{fig:attention-weights} illustrates three heat maps of self-attention weights on the last 15 positions, \textit{i.e.}, attention matrix $\mathbf{A}$ of the original STOSA/SASRec, attention matrix $\mathbf{A}$ of the reworked model using $\boldsymbol{M}_{simp}$, and attention matrix $\mathbf{B}$ refined by $\boldsymbol{M}_{simp}$, respectively. 
Compared to the original attention matrix of STOSA, the matrix $\mathbf{A}$ derived from our reworked model (\Cref{fig:attention-weights}(b)) tends to place significant emphasis on recent behaviors, where the values in the bottom right corner are relatively significant. This indicates that our mechanism enables the self-attention weights $\mathbf{A}$ to better grasp long-range dependencies through the training process. Moreover, the refined attention matrix $\mathbf{B}$ displays a more significant focus on recent behaviors and exhibits a certain regular attention pattern (as shown in \Cref{fig:attention-weights}(c)). This suggests that the refined matrix is more capable of capturing higher-order relationships. This pattern is also confirmed in the SASRec based methods, further verifying the validity of our proposed approaches. 

\begin{table*}[tp]
\centering
  \fontsize{8.4}{10.2}\selectfont
  \caption{NDCG@5 performance comparison \textit{w.r.t.} different attention layers on Beauty. The improvements (\%) of our mechanisms over the original method are denoted by the subscripts. Improvements over baselines are statistically significant with $p<0.01$.}
  \vspace{-2.5mm}
  \setlength{\tabcolsep}{2mm}{
  \label{tab:performance-layers}
    \begin{tabular}{c|cccc|ccccc}
    \toprule[1pt]
    \multirow{2}{*}{Layers}&\multicolumn{4}{c|}{\textbf{SASRec}}&\multicolumn{5}{c}{\textbf{STOSA}}\\  
&Original&$\boldsymbol{M}_{simp}$&$\boldsymbol{M}_{value}$&$\boldsymbol{M}_{add}$&Original&$\boldsymbol{M}_{simp}$&$\boldsymbol{M}_{value}$&$\boldsymbol{M}_{add}$&$\boldsymbol{M}_{stoc}$  \cr \hline
    1& 0.0234&0.0243$_{+3.85}$&0.0251$_{+7.26}$&0.0239$_{+2.14}$ &0.0285 &0.0304$_{+6.67}$ &0.0286$_{+0.35}$ &0.0305$_{+7.02}$ &0.0276$_{+3.16}$ \\
    2& 0.0242&0.0255$_{+5.37}$&0.0250$_{+3.31}$&0.0254$_{+4.96}$  
    &0.0271 &0.0286$_{+5.54}$ & 0.0278$_{+2.58}$&0.0289$_{+6.64}$ &0.0273$_{+0.74}$ \\
    3&
    0.0265&0.0274$_{+3.40}$&0.0270$_{+1.88}$&0.0278$_{+4.91}$
    &0.0257 &0.0265$_{+3.11}$ &0.0260$_{+1.17}$ &0.0277$_{+7.78}$ & 0.0261$_{+1.56}$\\
    4& 0.0247&0.0262$_{+6.07}$&0.0255$_{+3.24}$&0.0267$_{+8.10}$
    &0.0260 &0.0267$_{+2.69}$ & 0.0265$_{+1.92}$&0.0284$_{+9.23}$ &0.0270$_{+3.85}$ \\
    \bottomrule[1pt]
    \end{tabular}}
    \label{tab:layers}
    \vspace{-2mm}
\end{table*}

Our mechanisms can be seamlessly integrated into  single-layer self-attention (experiments in \Cref{tab:overall-performance} are achieved with one layer), and therefore we can repeat our reworked layer to multi-layer structure. Additionally, in \Cref{tab:layers}, we compare the performance of our mechanisms in multi-layer networks against original approaches on Beauty. 
The experimental results clearly indicate that our mechanism exhibits significant advantages even in the context of multi-layer self-attention. Despite the original models already achieving notable recommendation performance, our mechanisms can still further enhance their capabilities, providing additional improvements. 
This demonstrates the effectiveness and applicability of our mechanism in augmenting existing models, showcasing its potential for enhancing recommendation performance.
Among different mechanisms, $\boldsymbol{M}_{add}$ exhibits increasingly significant improvements as the number of layers expands. This further emphasizes the capability of $\boldsymbol{M}_{add}$ to balance attention weights in different contexts. 
The performance improvements of our mechanisms across different multi-head self-attention configurations are not explicitly compared, as results in \Cref{tab:overall-performance} are obtained by conducting a grid search on two original models to identify the best attention head configurations \textit{w.r.t.} different datasets at first. By deploying our mechanisms under the same configuration, we have demonstrated their effectiveness across different numbers of attention heads.

\begin{figure}
\centering
    \subfigure[SASRec]{
    \includegraphics[width=0.48\linewidth]{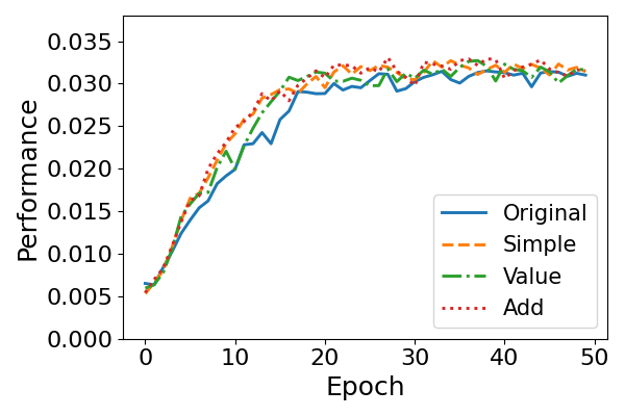}}
    \subfigure[STOSA]{
    \includegraphics[width=0.48\linewidth]{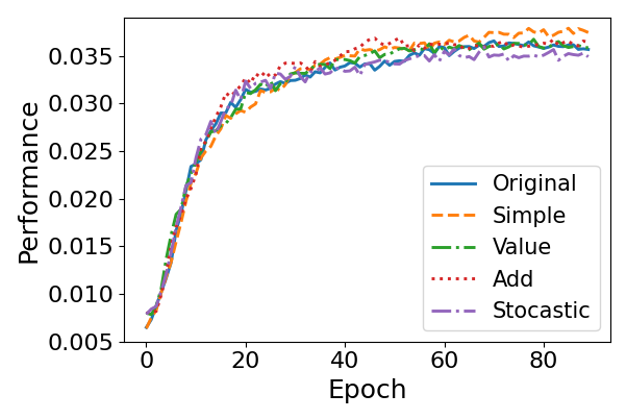}}
    \vspace{-2mm}
\caption{The validation performance (NDCG@5) trends 
using SASRec and STOSA on Beauty
.}
    \label{fig:NDCG-trends}
    \vspace{-3mm}
\end{figure}

\subsection{Qualitative Analysis}

In this section, we qualitatively visualize and analyze different models, which helps identify the significant difference between our mechanisms and original models. All comparisons are conducted under the same configuration. 

\Cref{fig:NDCG-trends} shows the validation performance trends of our mechanisms (integrated into SASRec and STOSA) on Beauty. Our methods, especially $\boldsymbol{M}_{simple}$ and $\boldsymbol{M}_{add}$, exhibit accelerated and enhanced learning trends. This suggests that our methods can efficiently improve recommendation performance by capturing higher-order correlations among items. Since the trends and results on other datasets follow a similar pattern, we have chosen not to present them individually.

\begin{figure}
\centering
    \subfigure[Beauty]{
    \includegraphics[width=0.48\linewidth]{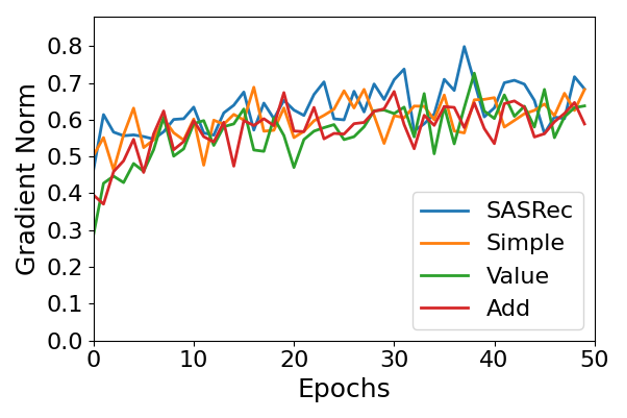}}
    \subfigure[ML-1M]{
    \includegraphics[width=0.48\linewidth]{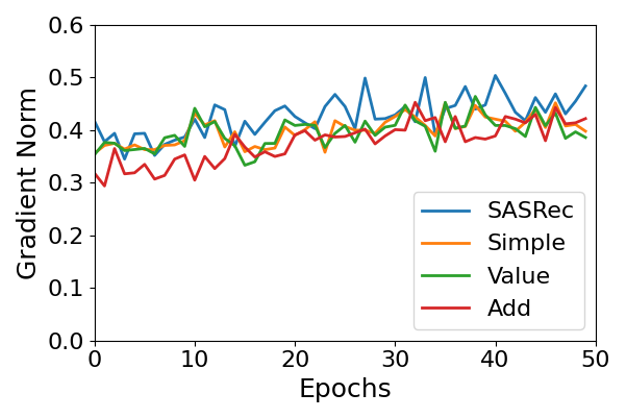}}
    \vspace{-2mm}
\caption{The global norm of gradients comparison using original SASRec. Best viewed in color.}
    \label{fig:norms}
    \vspace{-3mm}
\end{figure}

We use \Cref{fig:norms} to compare the global norm of gradients (final 50 epochs prior to early stopping) when employing different models on Beauty and ML-1M. We find that our mechanisms can accelerate the learning process by enabling the global gradient norm to reach a stable phase more quickly compared to the original model. This observation complements the conclusion drawn from \Cref{fig:NDCG-trends}, highlighting that the higher-order relationships captured by AWRSR facilitate the learning of sequential recommendation. \Cref{fig:NDCG-trends} shows that STOSA and SASRec share the similar learning trend, which is further supported in the gradient norm comparison. We therefore only present the results from different datasets in the context of SASRec.

%Theoretical analysis (mutual information \cite{cover1999elements}), complexity comparison, and additional experimental discussions (including global norm trends comparison and performance comparison w.r.t. different attention layers) can be found in the \textbf{auxiliary material}. 

\subsection{Complexity Analysis}
\label{sec-complexity}
%We analyze the space and time complexities that arise when our refinement mechanisms are implemented in the context of STOSA and SASRec. 
%The space complexities of the original SASRec and STOSA are $O\left(|\mathcal{V}| d+n d+d^2 / 2\right)$ and $O\left(|\mathcal{V}| d+n d+d^2\right)$, respectively. In terms of our mechanisms, the additional space complexity compared to the original architectures mainly comes from the need of storing new matrices. Taking $\boldsymbol{M}_{simp}$ as an example, we need to store $\mathbf{W}^{(RQ)}$, $\mathbf{W}^{(RQ)}$, and $\mathbf{B}$, each of which is $n \times n$, thus requiring extra $O(n^2)$ space in the context of both SASRec and STOSA, as the refinement is performed on the attention matrix of the same size. 
%This is moderate compared to original methods since it does not grow with the number of users, and $n$ (the maximum sequence length, and the usual practice is to set it according to the average length of users' interaction) is typically not big in recommendation problems. Similarly, the additional space complexity required by other mechanisms can also be easily calculated, which is typically not significant. 

We analyze the space and time complexities that arise when our refinement mechanisms are implemented in the context of STOSA and SASRec. 
The space complexities of the original SASRec and STOSA are $O\left(|\mathcal{V}| d+n d+d^2 / 2\right)$ and $O\left(|\mathcal{V}| d+n d+d^2\right)$, respectively. In terms of our mechanisms, the additional space complexity compared to the original architectures mainly comes from the need of storing new matrices. Taking $\boldsymbol{M}_{simp}$ as an example, we need to store $\mathbf{W}^{(RQ)}$, $\mathbf{W}^{(RQ)}$, and $\mathbf{B}$, each of which is $n \times n$, thus requiring extra $O(n^2)$ space in the context of both SASRec and STOSA, as the refinement is performed on the attention matrix of the same size. 
This is moderate compared to original methods since it does not grow with the number of users, and $n$ (the maximum sequence length, and the usual practice is to set it according to the average length of users' interaction) is typically not big in recommendation problems. Similarly, the additional space complexity required by other mechanisms can also be easily calculated, which is typically not significant. 

Moving forward, we analyze the time complexity of the original models, with SASRec having $O\left(n d+n^2 d+4 n^2+\frac{n d^2}{2}\right)$ and STOSA having $O\left(n^2 d+n d^2\right)$. Note that these time complexities are typically dominated by the self-attention layer, as analyzed in \cite{fan2022sequential}. As for our mechanism $\boldsymbol{M}_{simp}$, we need to apply matrix multiplications to transforming attention weights $\textbf{A}$ and dot product for refining weights, contributing the additional time complexity $O\left(n^2n\right)$. In practical experiments on Beauty and Anime datasets (described in Experiments), the best performance (for both AWRSR in the context of SASRec and original SASRec) is achieved with the settings ($d=128, n=20$) and ($d=128, n=30$), respectively, \textit{i.e.}, $d$ is usually much larger than $n$. This means that the dominant term of our mechanisms still be from the original self-attention layer. Overall, the proposed mechanisms are not computing a new self-attention layer. Instead, we reuse the attention weights from the current layer and adjust them with trainable weights. They are less computationally demanding than calculating a full new self-attention layer.

\vspace{-1mm}
\section{Related Work}
\vspace{-0.5mm}

%In this section, we review the literature relevant to our research in three primary areas. 

\textbf{Sequential Recommendation} 
Sequential recommendation significantly aligns with real-world application scenarios in recommender systems, as it aims to predict the next item that users might be interested in, based on their historical interactions or preferences. Early methods in sequential recommendation mainly focus on sequence processing techniques such as Markov Chains \cite{shani2005mdp, he2016fusing} and sequence-aware collaborative filtering \cite{liu2010adapting, zhao2013interactive}. However, these approaches often struggled to capture complex patterns and long-range dependencies in user-item interaction sequences.
To address these limitations, researchers have started to explore deep learning techniques for sequential recommendation. Specifically, models of Recurrent Neural Networks (RNNs) \cite{hidasi2015session, hochreiter1997long}, Long Short-Term Memory (LSTM) \cite{devooght2017long}, and Gated Recurrent Units (GRUs) \cite{cho2014learning} have been widely adopted to capture temporal dependencies in user-item interactions. Motivated by capturing nuanced relationships among items in sequential data, some studies try to extend item-item relationships to higher-order item transitions, like Markov Chains-based Fossil \cite{he2016fusing} and distribution representations-based DVNE \cite{zhu2018deep}.
\\
\textbf{Self-attention in SR}. 
SA mechanism, as a fundamental component of the Transformer \cite{vaswani2017attention}, has emerged as a primary technique to capture more complex relationships and dependencies within user-item interaction sequences. The primary advantage of self-attention in this context is its ability to capture long-range dependencies and complex patterns in user-item interactions, which is a known limitation of traditional RNN and LSTM-based models \cite{kang2018self}.
Recent studies have explored the integration of SA mechanisms in sequential recommendation.
For instance, the SASRec model \cite{kang2018self} employs SA mechanism to capture dependency patterns in user-item interaction sequences. Similarly, the BERT4Rec model \cite{sun2019bert4rec} fine-tunes a pre-trained BERT model for the recommendation task, demonstrating the effectiveness of self-attention technique in SR. The NARM model \cite{li2017neural} employs a combination of LSTM and SA to capture both local and global patterns in action sequences.
These studies have demonstrated the effectiveness of self-attention mechanisms in sequential recommendation, outperforming traditional methods in various benchmark datasets. 
\\
\textbf{Self-Attention Weights}.  
One critical aspect of the SA mechanism is the attention weights, whose importance lie in their ability to focus on different levels of importance within the sequence.  
Several studies have focused mainly on designing different attention structures to better capture relationships in sequential data, including
local attention \cite{guo2019progressive},
global-local attention \cite{li2017image}, and
hierarchical attention \cite{ying2018sequential}.
These approaches recognize the significance of attention mechanisms in modeling dependencies and have sought to enhance their effectiveness through attention weight designs.
Furthermore, numerous studies have explored the interpretation of attention mechanisms by analyzing the distribution of attention weights, such as
analyzing both trained and untrained attention metrics \cite{zhong2019fine}, proposing alternative tests to determine when/whether attention can be used as an explanation \cite{jain2019attention, wiegreffe2019attention}, and designing
graph reasoning tasks to study attention in a controlled environment \cite{knyazev2019understanding}. 
Recently, a novel approach called STOSA (Stochastic Self-Attention) \cite{fan2022sequential} has been proposed, introducing a fresh perspective on measuring attention weights, which measures attention weight as the negative 2-Wasserstein distance between items' stochastic representations. 
Through our review of existing self-attention studies, we can identify a gap in current research. It appears that no work to date has investigated the correlations that may exist among the attention weights themselves.

\vspace{-1mm}
\section{Conclusion}
\vspace{-0.5mm}
This study offers a novel perspective on improving sequential recommendation by refining the attention weights in self-attention mechanisms. We develop a specific strategy that involves "paying attention to attention", with the goal of revealing and leveraging the higher-order transitions hidden in attention weights. This is accomplished by utilizing two distinct and representative self-attention models (\textit{i.e.}, SASRec and STOSA), to obtain different attention weights and design the corresponding refinement mechanisms.
The improvement on SASRec (a seminal self-attention-based recommendation model) underlines our method's wide applicability. Meanwhile, the enhancement seen with STOSA (a novel model with a distinctive attention weight computation) showcases the robustness and flexibility of the proposed refinement mechanisms. 
Our findings provide valuable insights for future research, particularly in exploring alternative methods of refining attention weights. The refinement perspective extends beyond the SR field and has the potential for application in other domains that leverage SA mechanisms.

\vspace{-1mm}
\section{Acknowledgments}
\vspace{-0.5mm}
 This work is supported in part by the National Natural Science Foundation of China under Grant 61862053, and by the Qinghai Provincial Department of Science and Technology Applied Basic Research Project under Grant 2021-ZJ-717. This work is supported by high performance computing center of Qinghai University.

%%
%% The next two lines define the bibliography style to be used, and
%% the bibliography file.
\bibliographystyle{ACM-Reference-Format}
\bibliography{sample-base}

%%
%% If your work has an appendix, this is the place to put it.

\end{document}